# LIQUID JET IN CROSSFLOW: REVIEW OF BREAKUP MODES & INJECTOR GEOMETRY EFFECTS

**Anubhav Sinha**

Department of Mechanical Engineering, Indian Institute of Technology (Banaras Hindu University), Varanasi, India
er.anubhav@gmail.com

## Abstract

This review focuses on the liquid jet in crossflow (LJIC) configuration. LJIC is one of the most common strategies used for fuel injection in aerospace applications. It is popular due to its simplicity and efficient atomization characteristics. The aerodynamic force of the airflow is utilized to break liquid jet into small droplets. The objective of the present work is to give a basic overview of the physical processes involved in the breakup and penetration of LJIC. Breakup modes and underlying mechanisms are discussed in detail. Various modes are described and associated non-dimensional numbers are explained. Injector geometry which is often overlooked in literature is paid special attention. The mechanism of liquid jet instability getting triggered by velocity profile redistribution is explained using experimental and computational results. Surface waves on liquid jets are discussed. A theoretical model used to predict the wavelength of surface waves is described. DNS results are used to demonstrate the growth of surface instability on a liquid jet. Jet penetration and trajectory in the presence of crossflow are discussed. Various trajectory equations and the parameters used are discussed in detail. Progress in computational studies for LJIC is highlighted and challenges are discussed.

## Nomenclature

| | |
|---|---|
| $\rho$ | Density |
| $V$ | Velocity |
| $D$ | Jet diameter/ length scale |
| σ | Surface tension |
| μ | Viscosity |
| $q$ | Momentum flux ratio or momentum ratio |
| $Re$ | Reynolds number |
| $We$ | Weber number |
| $Oh$ | Ohnesorge number |
| $\eta$ | Wave growth rate |
| $\lambda$ | Wavelength |
| $k$ | Wave number |
| $a$ | acceleration |
| $\theta$ | momentum boundary layer thickness |

### Subscripts

| | |
|---|---|
| g | gas |
| l | liquid |

### Abbreviations

| | |
|---|---|
| LJIC | Liquid Jet in Crossflow |
| DNS | Direct Numerical Simulations |
| LES | Large Eddy Simulations |

## 1. INTRODUCTION

Jet in crossflow (JIC) or transverse jet is a common configuration for various applications, both in natural processes and in engineering devices [1, 2]. A smoke plume from a chimney getting deflected due to wind is a common example of a jet in crossflow. JIC is a prominent strategy for fuel injection, particularly in aerospace applications [3-5]. This configuration is used in gas turbine combustors, afterburners, ramjets, scramjets, etc. A liquid jet is injected transverse to the crossflow, and the incoming air momentum is used to break the jet into smaller ligaments and droplets which eventually evaporate and mix. If the jet breaks too quickly, then it will not be able to penetrate and mix with the main flow. On the other hand, if it breaks too late, the mixing will not be complete until the combustion stage and combustion efficiency will be decreased. Hence, an optimum design will consider these scenarios and attempt to balance between these two extremes. Owing to the significance of this configuration, there is a plethora of literature on this subject, as summarized in previous review articles [6-8]. Many experimental studies have been carried out for a wide range of parameters. For experimental studies, a wind tunnel is generally used to generate air flow, and the jet nozzle is flush mounted to the test section wall injecting liquid perpendicular to the airflow. The test section has transparent walls, and the jet structure is imaged using various imaging techniques like pulsed shadowgraphy, holography, backlit imaging, etc. A bright light source such as a laser and high-resolution cameras has aided in our understanding of jet breakup morphology. High-speed images are required to understand the sequence of breakup steps. Instantaneous and averaged flow field images are used to study jet trajectory, and penetration and characterize breakup behaviour. Flow parameters are identified which impact resultant breakup and fuel mixing. Experimental studies have mostly reported atmospheric conditions, which are easier to simulate in a lab-scale facility. Some studies also report higher pressure and temperature which are closer to the actual working conditions for aerospace applications. However, as the parameters are reported in non-dimensional numbers, results from atmospheric conditions can be used to explain the underlying physics and estimate jet behaviour in higher pressures and temperatures. Fig. 1 shows some typical examples of LJIC cases highlighting different degrees of atomization.

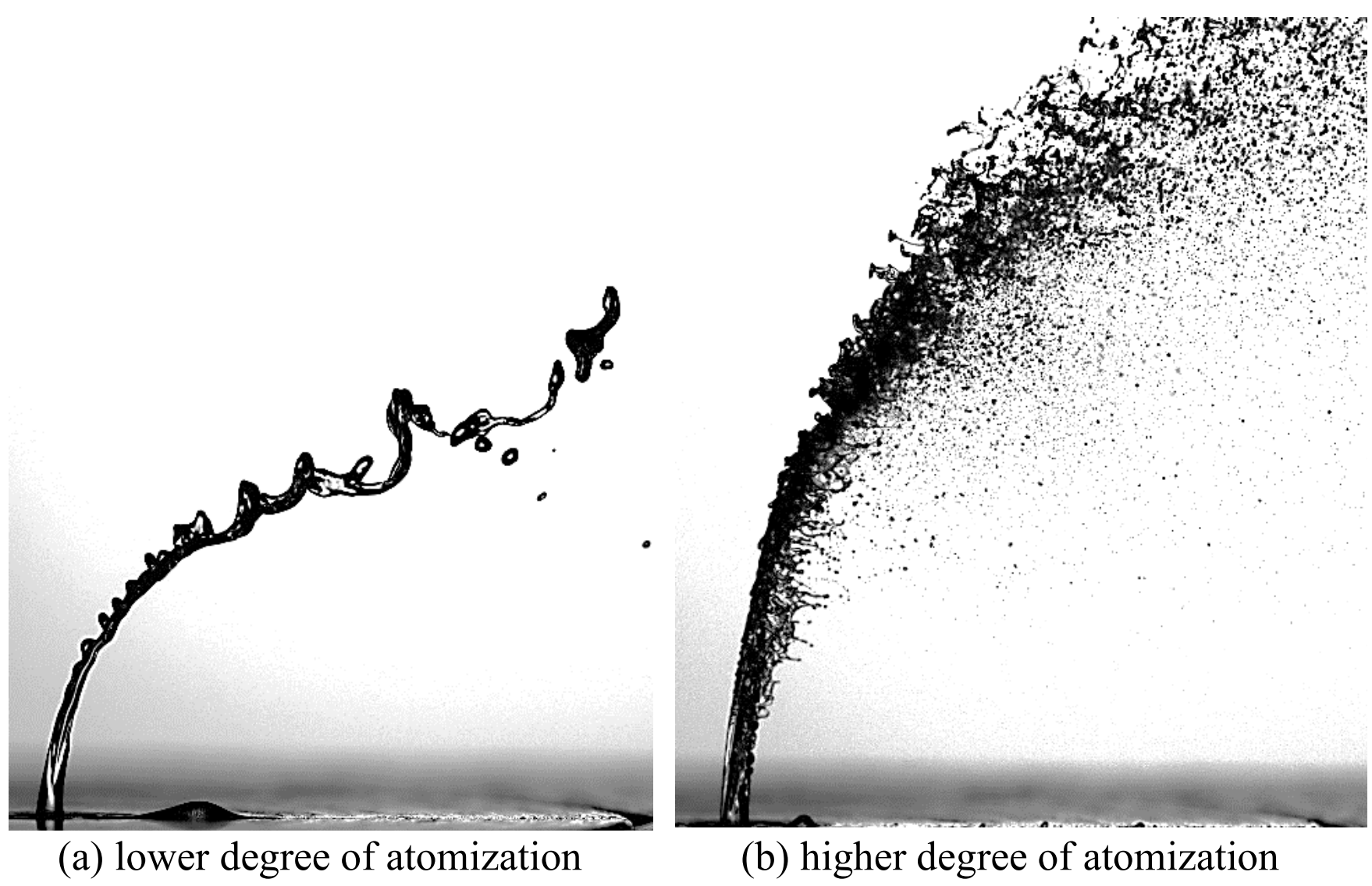

(a) lower degree of atomization (b) higher degree of atomization

**Fig. 1.** Liquid jet in crossflow – some typical examples

The liquid is injected vertically from the bottom wall while air is coming from the left side in the view shown in Fig. 1. Figure 1(a) presents a lower degree of atomization while Fig. 1(b) demonstrates a more disruptive or higher degree of atomization. How we classify these cases and the factors that control this behaviour will be discussed in detail in later sections.
Modeling of LJIC configuration is one of the most challenging problems in CFD. There are various complexities involved such as modeling multiphase flow, capturing the liquid-gas interface accurately, capturing primary and secondary breakup, etc. Further, modeling for two phase flows becomes much complicated due to the density difference between the liquid and gas phases which makes the computational matrix stiff. There are various computational modeling approaches employed to model this configuration and is an active area of research. However, most of the reported literature on computational modeling has used higher ambient pressure to reduce the density ratio. Density ratio is the ratio of liquid and gas density. For water in ambient air at atmospheric pressure, the density ratio is around 800. This poses a significant challenge for numerical modeling as sharp density jumps can lead to numerical stability issues. Hence, most of the reported work on LJIC is on higher ambient pressure. Higher pressure is also closer to actual conditions, but the lack of experiments at higher pressures makes it difficult to validate those results. Recent advances in DNS studies have helped to gain some insights, but still, much remains to be done in this regard.
This chapter focuses on the primary breakup of round liquid jet injected perpendicular to airflow. Supersonic flows, evaporating conditions, etc. are not the focus of this discussion. The objective is not to compile a summary of the available literature, but to provide an overview of various processes involved and underlying mechanism of breakup in LJIC. In section 2, breakup modes of LJIC are explained and regime maps proposed by various groups are discussed. Further, in section 3, the significance of the role of injector geometry is highlighted presenting some recent results. Jet penetration and various correlations to estimate jet trajectory are presented in section 4. Section 5 focuses on the computational modeling aspects highlighting various modelling approaches and challenges involved. The chapter ends with a summary and discussion of future research in this field.

## 2. BREAKUP MODES

The impact force from the momentum of air is primarily responsible for the jet breakup, especially in the case of laminar jets. On the other hand, the surface tension of the liquid opposes breakup and tries to stabilize the jet surface. The balance between these two opposing forces can be quantified in terms of non-dimensional Weber number ($We_g$) which is the ratio of inertial and surface tension forces and can be expressed as:

$$We_g = \frac{Inertia}{Surface\ tension} = \frac{\rho_g V_g^2 D}{\sigma} \tag{1}$$

where $\rho_g$ is the gas density, $V_g$ is the crossflow velocity, $D$ is the length scale, generally taken to be the jet diameter, and $\sigma$ is the surface tension. Subscript g in Weber number indicates that is an aerodynamic Weber number, which is based on gas properties. Weber number based on liquid properties can also be defined, but it is useful for liquid jet studies in quiescent ambient, not in the presence of crossflow. The aerodynamic Weber number (henceforth referred to as Weber number for brevity) is one of the dominant factors governing the breakup behavior of liquid jets in crossflow. Jet breakup in LJIC is quite similar to the breakup modes exhibited in the secondary breakup of droplets. One of the earliest regime maps for jet breakup in LJIC was proposed by Wu et al. [9]. This regime map does not consider the effect of liquid turbulence, cavitation, and effervescence. The map shown considers parametric space defined by Weber number (*We*) and jet to crossflow momentum flux ratio (*q*). Momentum flux ratio or simply momentum ratio (*q*) can be defined as:

$$q = \frac{Liquid\ jet\ momentun}{Air\ momentum} = \frac{\rho_l V_l^2}{\rho_g V_g^2} \tag{2}$$

where $\rho_l$ is the liquid density, $V_l$ is the liquid jet velocity, $\rho_g$ is the gas density, and $V_g$ is the crossflow air velocity. While Wu et al. [9] consider *q* and *We* to be the dominant parameters, Sallam et al. have taken *Oh* and *We* parametric space. *Oh* is the Ohnesorge number, which can be defined as:

$$Oh = \frac{Viscous}{Interia\ \cdot\ Surfacer\ tension} = \frac{\mu}{\sqrt{\rho\ \sigma\ D}} \tag{3}$$

where $\mu$ is the liquid viscosity, $\rho$ is the liquid density, $\sigma$ is the surface tension and *D* is the jet diameter. *Oh* can also be expressed in terms of Weber number (*We)* and Reynolds number (*Re*):

$$Oh = \frac{\sqrt{We}}{Re} \tag{4}$$

The regime map proposed by Aalburg et al. [10] shows dependence on Oh, but for small Ohnesorge numbers (which is usually the case), *We* is the deciding factor, which agrees with Wu et al. [9]. The breakup behaviour of LJIC can broadly be divided into two major types- ***column breakup*** and ***surface breakup*** [9]. Figure 2 depicts this behaviour in LJIC instantaneous images with a magnified view in the near nozzle region. Fig. 2(a) shows the column breakup while 2(b) shows the surface breakup case. The same cases were presented in Fig. 1, where the column breakup case falls in a lower degree of atomization and surface breakup exhibits a higher degree of atomization. The column breakup case shows bending and flattening of the column, which is visible in the near-nozzle region. However, no droplets or ligaments are produced before the column fracture point. Whereas, in surface breakup conditions, small droplets are being stripped from the lateral side of the jet even at very close to the nozzle exit. The droplets produced in the surface breakup case appear to be much smaller than those produced by the column breakup (cf. Fig. 1). As mentioned above, the classification proposed by Wu et al. [9] assumes a laminar jet. However, a turbulent jet image exhibited the characteristic features of surface stripping and is used in Fig. 2.

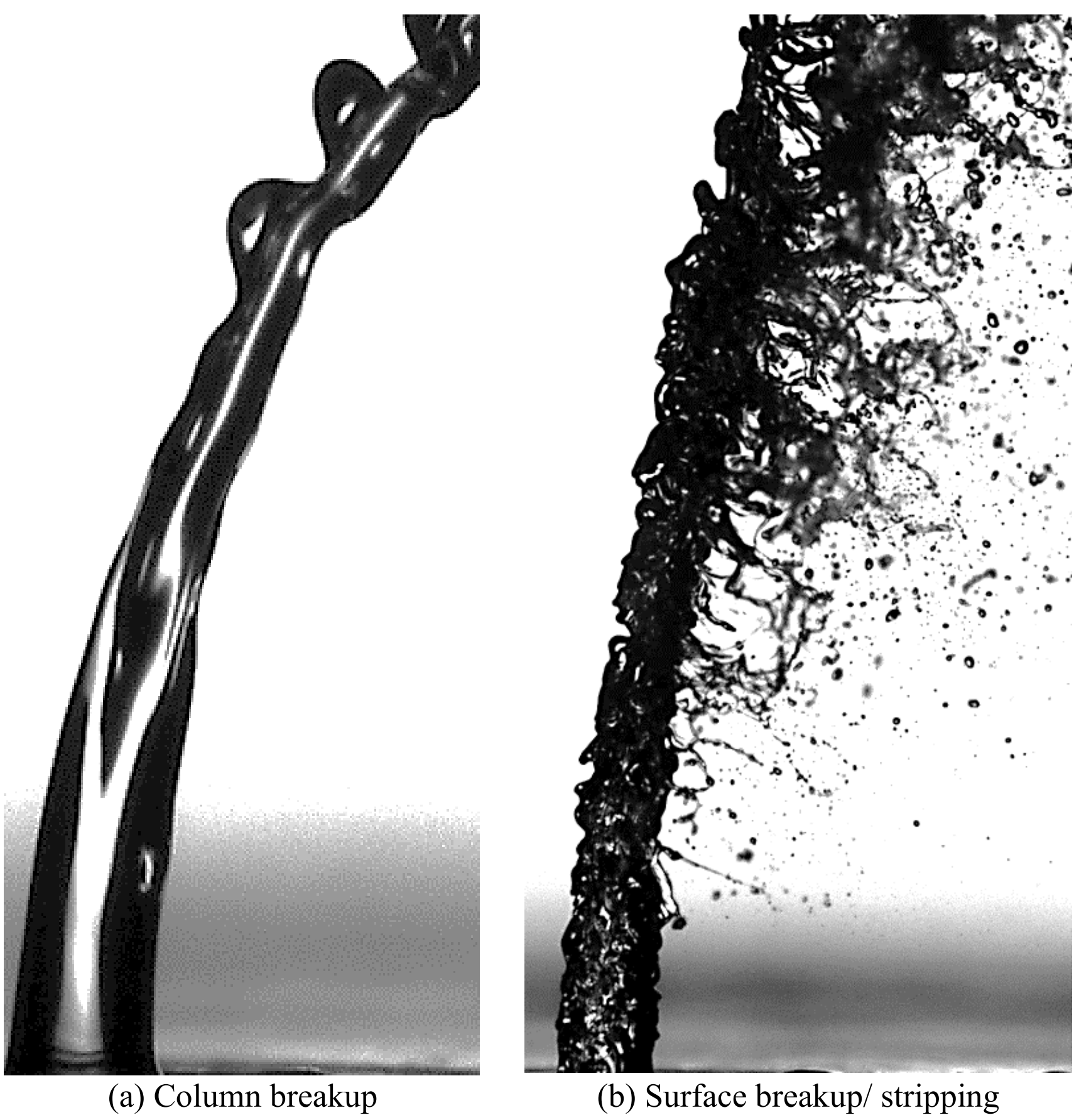

(a) Column breakup (b) Surface breakup/ stripping

**Fig. 2.** LJIC - Jet breakup modes – magnified views

LJIC breakup can further be divided into these regimes as per Wu et al. [9]:

1) Enhanced capillary breakup
2) Bag breakup
3) Multimode breakup and
4) Shear breakup.

The list presents regimes in order of increasing Weber's number. These regimes will be discussed in detail in the following paragraphs.

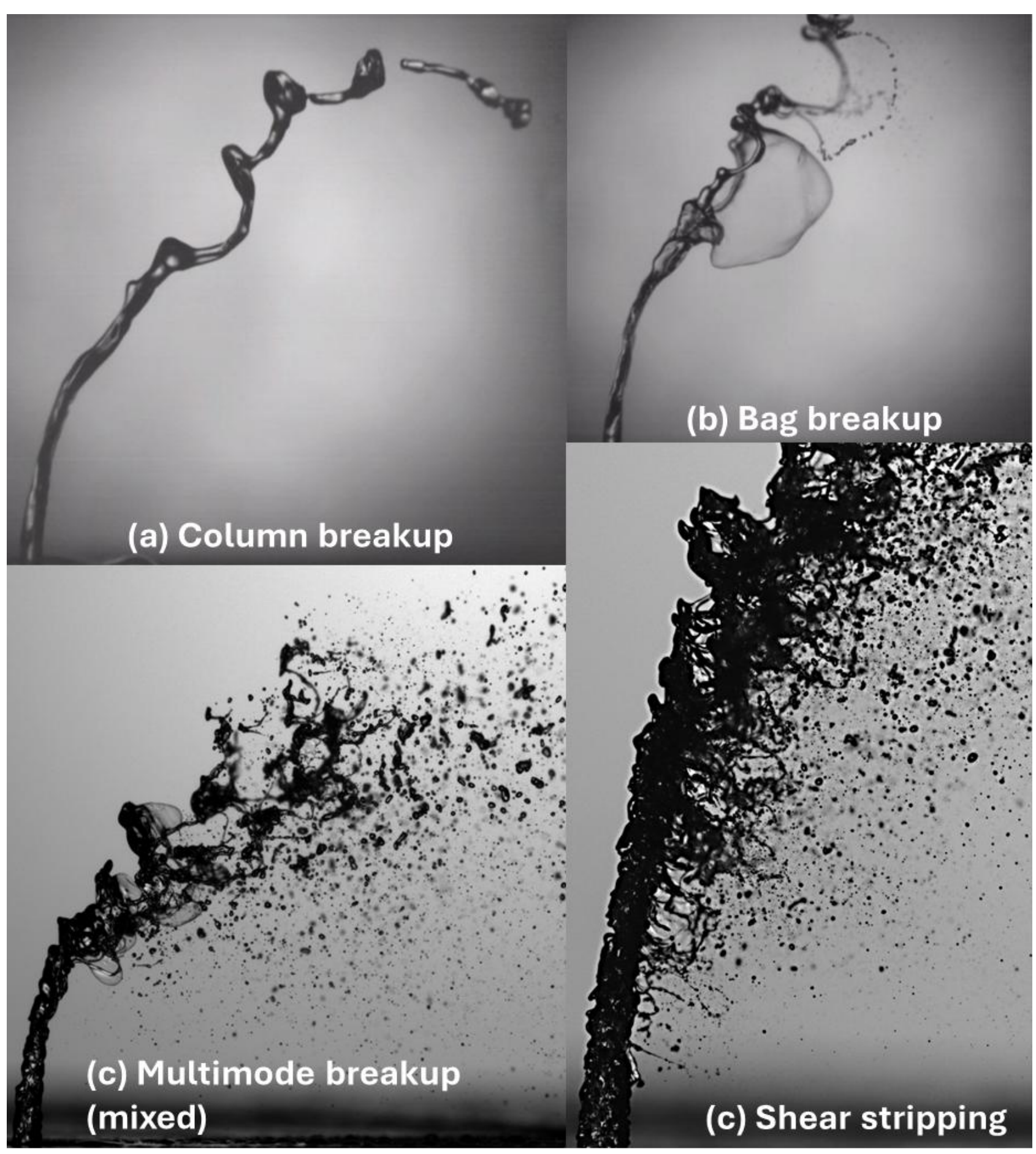

**Fig. 3.** LJIC- Various breakup modes

**1) Enhanced capillary breakup -** This regime has the lowest Weber number and hence the weakest aerodynamic forces. The jet breaks due to its inherent instability. It is similar to Rayleigh/ capillary breakup with some aid from the air momentum. The jet flattens to some extent, which increases the aerodynamic drag and bends the jet in the direction of the crossflow. However, the aerodynamic forces are not strong enough to form a bag, as observed in the next regime. The jet sections get stretched, forming lobes. The neck region between lobes gets thinner with time and eventually breaks, as observed near the top right corner.

**2) Bag breakup**– This is also a low Weber number mode. In this regime, the jet

gets flattened at the centre in the shape of a bag. First, a region in a jet flattens and takes a bag shape, then this bag grows and finally breaks. Various stages of bag formation are- bag inception, bag growth, and finally bag rupture. The bag is flattened to a very small thickness which eventually disintegrates to produce small droplets. The bag rim breaks and produces larger droplets. These two different sets of droplets produce a bimodal droplet size distribution in the near nozzle region. The droplet size produced by the bag is much smaller than that produced by the bag rim.

**3) Multimode breakup** – This mode is intermediate between Bag breakup and Shear breakup and shows mixed behaviour exhibiting characteristics of both bag and shear breakup. This model is shown in Fig. 3(c). A close examination in the near nozzle region shows bags being formed. Whereas, at higher locations, the jet is being sheared. This mode produces small droplet sizes as compared to previous regimes.

**3) Shear breakup** –This is the highest Weber number mode. Here the aerodynamic forces are so strong that they strip droplets from the lateral side of the jet. This is shown in Fig. 3(d), and also in Fig. 1(b). This is like shear or catastrophic breakup reported in droplets. This mechanism produces the smallest droplet sizes. A more violent form of the breakup can be seen in Fig. 2(b). The above classification is done assuming a laminar liquid jet. For a turbulent jet, another regime map is proposed by Sallam et al. [11]. The parameters used are liquid Reynolds number (*Re*) and a new parameter called Faeth number (*Fa*) defined as:

$$Fa = We_l \, q^{1/3} \tag{5}$$

Where $q$ is the momentum ratio, and $We_l$ is the liquid Weber number, which can be expressed as:

$$We_l = \frac{\rho_l V_l^2 D}{\sigma} \tag{5}$$

Sallam et al. [11] have carried out an experimental study of laminar and turbulent liquid jets in crossflow. They observed that below Fa=17000, the breakup can be classified as an Aerodynamic breakup, where the air momentum controls breakup behaviour. However, for higher Faeth numbers, liquid turbulence becomes the dominant parameter. They also reported the formation of ligaments at the upwind side of the jet surface for a turbulent jet, which is not observed for laminar jets. Interestingly, for higher jet turbulence levels, increasing the crossflow velocity can lead to suppression of upwind breakup, which is counter-intuitive but can be explained. As the breakup is caused primarily by the jet turbulence, higher air velocity suppresses it. There are some recent efforts to formulate the regime map in terms of Weber number and liquid Reynolds number [12, 13]. It can be argued that the air momentum in $q$ is accounted for in the Weber number and jet momentum gets reflected in liquid Reynolds number. Moreover, the *Re* can also be additionally used to estimate whether the jet is laminar or turbulent.
While Madabhushi et al. [13] fix Re=5000 to be the boundary between surface and column breakup, Sinha [12] has also reported the dependence on Weber number in regime boundaries, in addition to *Re*. Fig. 4 shows the combined dataset from the experiments of Sinha [12] and Madabhushi et al. [13] in the form of a regime map. The jet images are classified into column, surface, and mixed type of breakup.

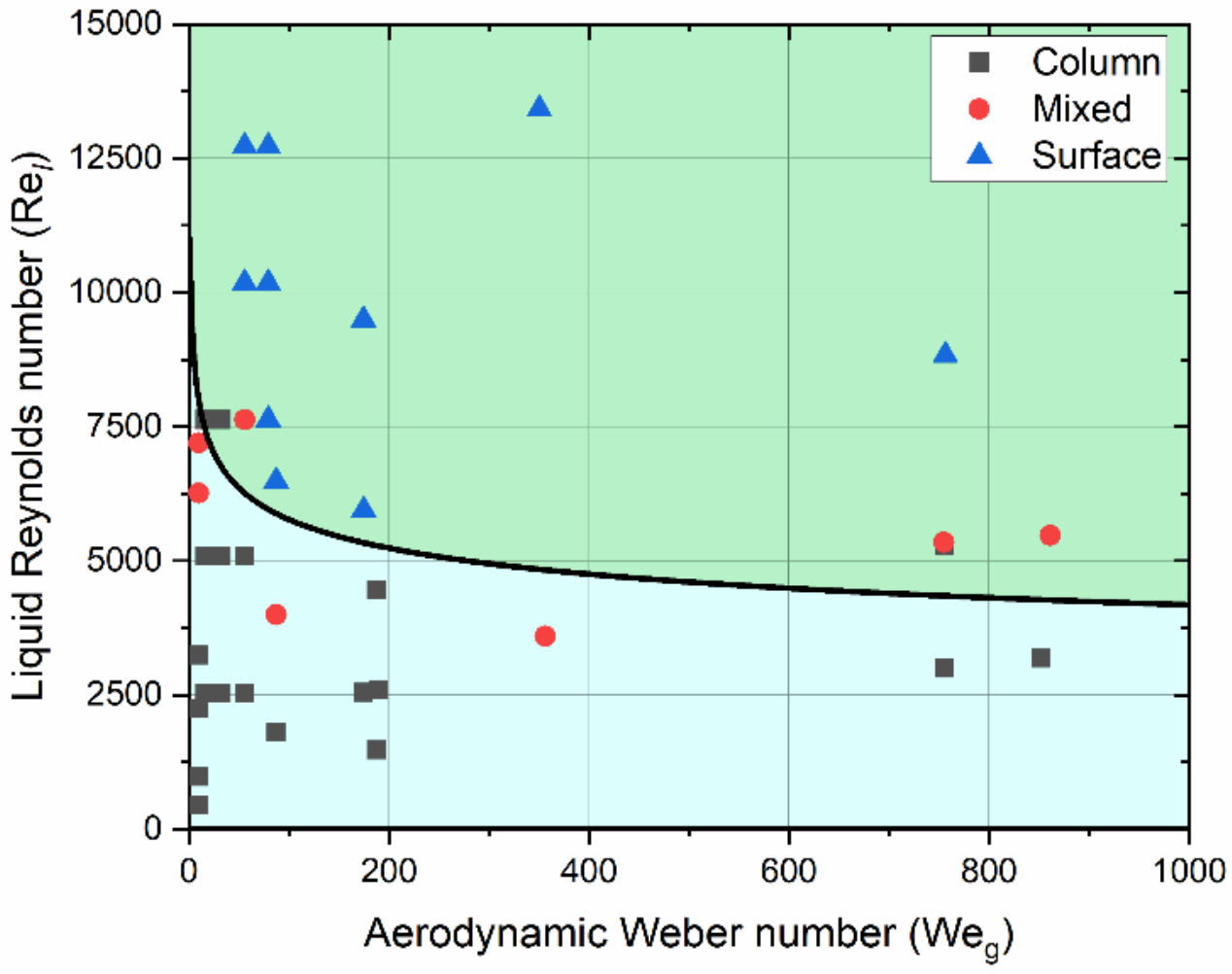

**Fig. 4.** Regime map with data combined from Sinha [12] and Madabhushi et al. [13].

It is important to understand that the regime boundaries are an estimate and sometimes the distinction between various regimes is not very clear and is subjective. Hence, it becomes important to analyse multiple instantaneous images, videos, and even advanced post-processing techniques like Proper Orthogonal Decomposition (POD) to identify modes of jet breakup [14, 15]. Further, there will be differences due to experimental error, optical method used, image processing, etc. However, one major factor that can significantly influence the jet behaviour and is often overlooked in injector geometry, which will be discussed in detail in the next section.

## 3. EFFECT OF INJECTOR GEOMETRY

Injector geometry determines the nature of the liquid jet. In a simple experiment, where the length-to-diameter ratio (*L/D*) of the injector was varied in LJIC experiments [15], the morphology of the jet was observed to change drastically. Fig. 5 shows two sets of images magnified in the near nozzle region. Fig. 5(a) and Fig. 5 (b) are at the same operating conditions, with a jet Reynolds number of 5090, while Fig. 5(a) looks laminar and glassy, and Fig. 5(b) is clearly turbulent. Similarly, Fig.

5(c) appears to be in a transition phase, with a slightly clear appearance on the windward side, and wave formation on the leeward side. However, Fig. 5 (d) demonstrates much higher turbulence levels and shows droplet stripping, despite the same Weber number as in Fig. 5(c). This figure clearly demonstrates that changing the nozzle design significantly affects the jet behaviour, even influencing the laminar to turbulent transition.

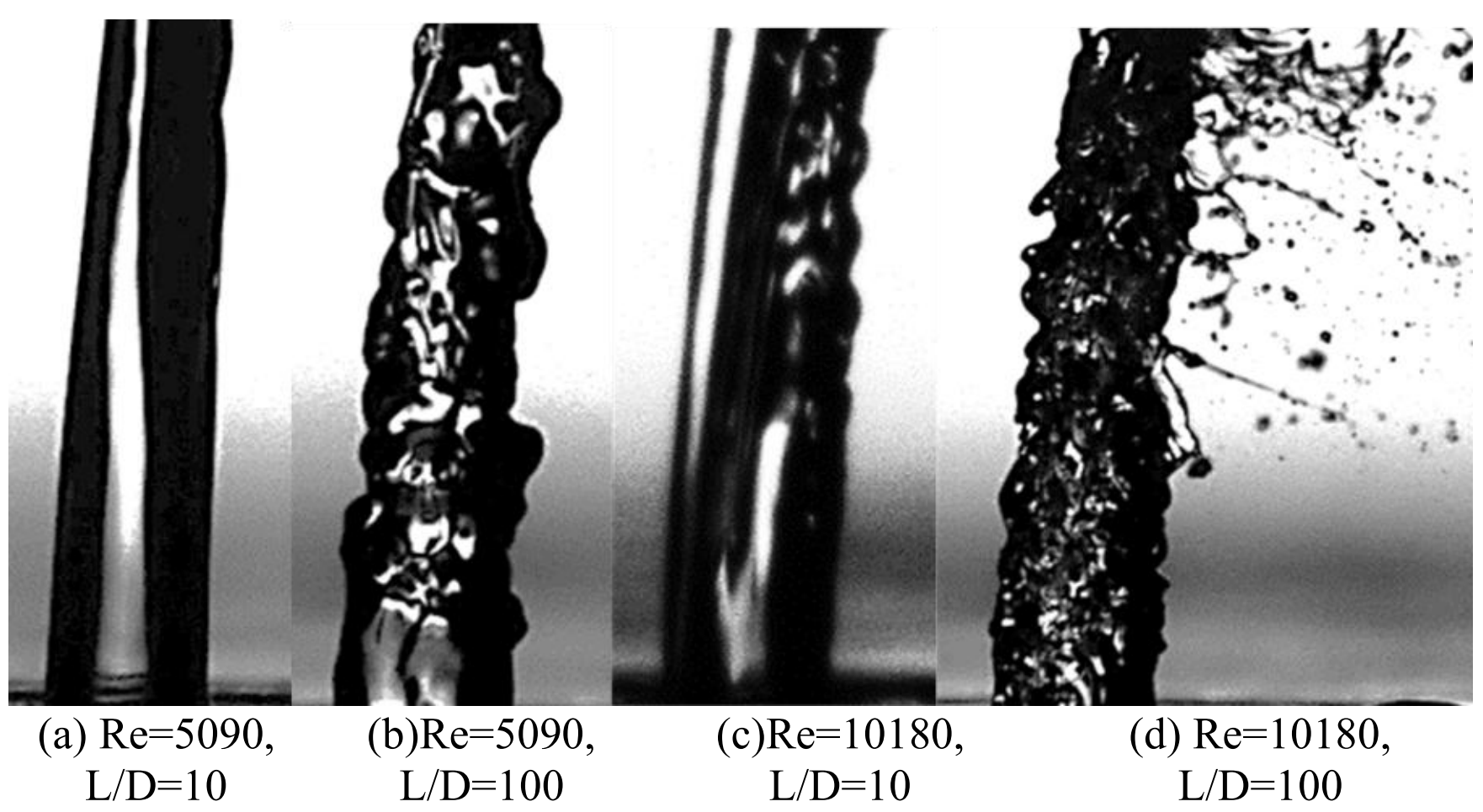

(a) Re=5090, L/D=10 (b)Re=5090, L/D=100 (c)Re=10180, L/D=10 (d) Re=10180, L/D=100

**Fig. 5.** Effect of injector geometry on jet LJIC (adapted from Sinha [15])

Prakash et al. [14] measured the height of the column breakup point (CBP) for two injectors with L/D ratios of 10 and 100. Their measurements demonstrate that the breakup point for the shorter nozzle (L/D=10) is higher than the longer nozzle (L/D=100). This is similar to the observation in Fig. 5 that a longer nozzle is more likely to give a turbulent jet that breaks earlier. Sinha [21] experimentally investigated the effect of injector geometry on jet behavior and summarized the findings in a regime map. The effect of the L/D ratio on laminar to turbulent transition for an LJIC configuration is captured in that regime map. It demonstrates that laminar to turbulent transition is not only a function of liquid Reynolds number but also influenced by injector geometry.

To isolate any influence from the crossflow air, straight liquid jets were also examined for the same set of injectors [14, 15]. It is reported that the jet for L/D of 10 is laminar at the injector exit. Still, it develops instability at some distance from the exit and becomes turbulent. As the L/D ratio is increased, the point of transition shifts close to the injector exit, and for L/D 50 and 100, the fully turbulent jet is observed at the exit. The question is why such a minor change in injector design (L/D) has such a profound influence on the resultant jet structure and breakup behavior. This can be understood by considering what happens to a liquid jet as it leaves the injector. As shown in the schematic in Fig. 6, there is a no-slip condition imposed on the liquid at the injector wall when the liquid is inside the injector. As soon as the liquid comes out, the no-slip condition is relaxed and there is a velocity profile redistribution between the liquid layers.

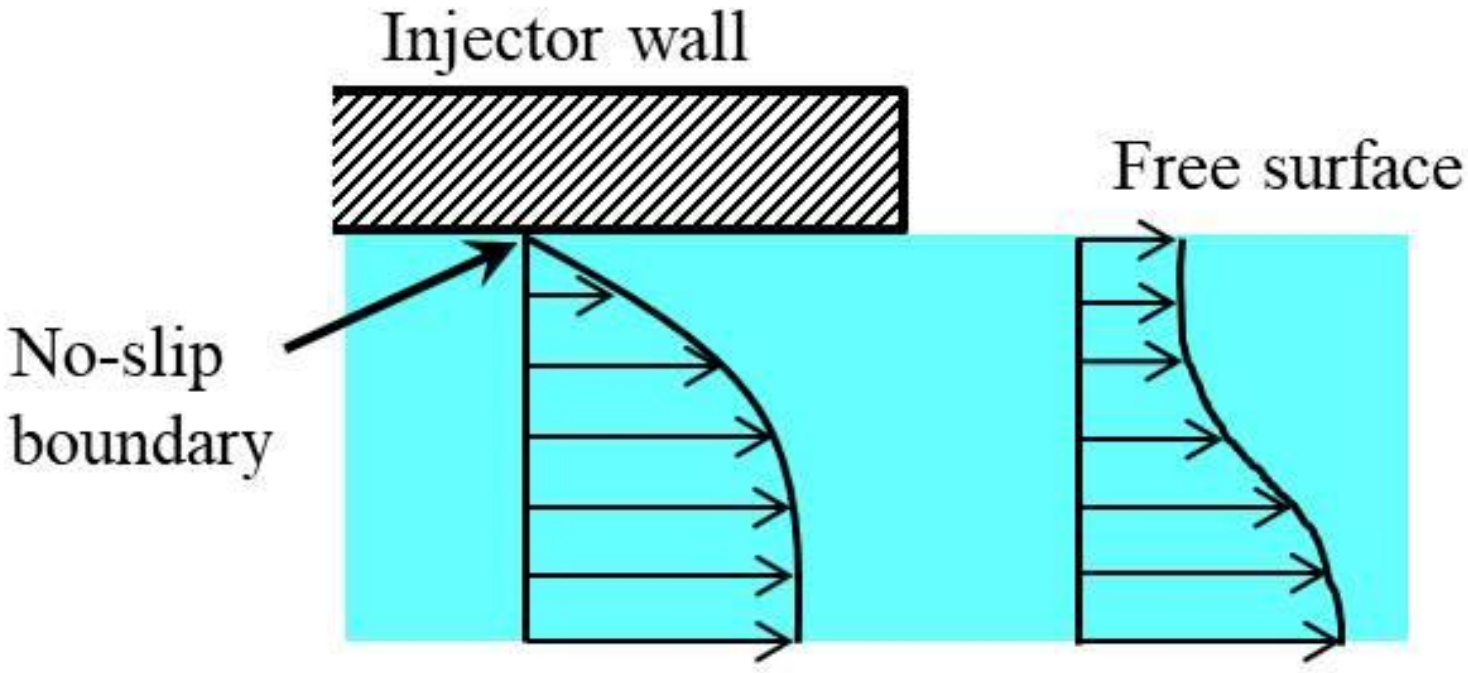

**Fig. 6.** Velocity profile redistribution near the wall at the injector exit**.**

This redistribution creates perturbation near the jet surface which triggers instability and eventually turbulent behavior. This hypothesis is further verified by recent DNS results (Srinivasan and Sinha [16]). Fig. 7 shows an instantaneous image of a straight liquid jet (without crossflow). Although the operating conditions for this case are quite different from the experimental study, the flow configuration and basic physics are the same. Similar to experimental observation, the liquid comes out from the injector as a smooth and laminar jet, but instabilities soon appear and render the jet turbulent. The image is colored by the liquid axial velocity. As evident, the surface velocity of the jet at the injector exit is zero. However, it begins to increase and reaches close to the peak velocity near the jet tip. It can be inferred that the velocity profile redistribution is causing instability which triggers turbulence breakup. Wave formation in LJIC will be explored in the next section.

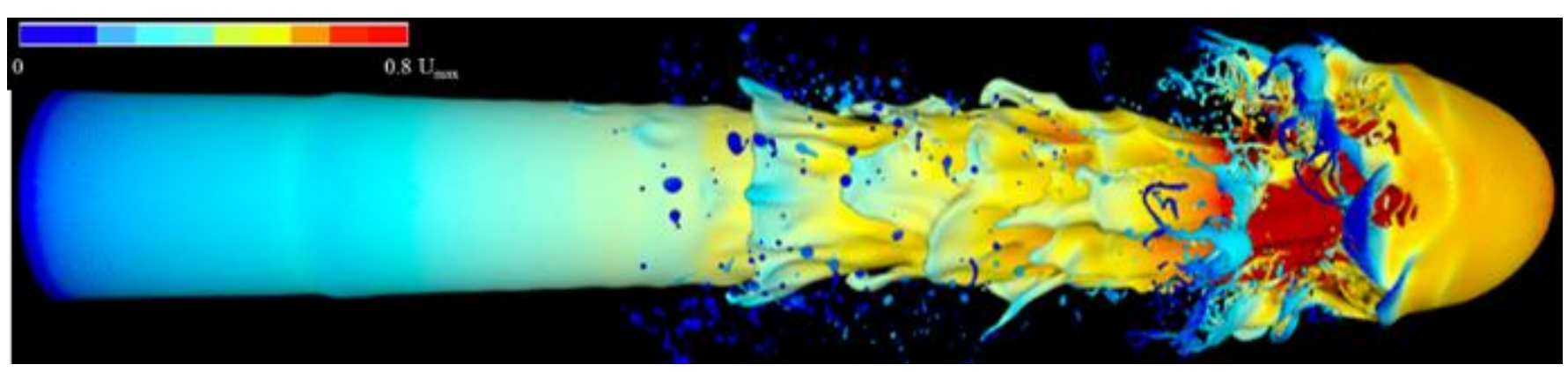

**Fig. 7.** Velocity profile redistribution at the surface of liquid jet**.**

## 4. SURFACE WAVES AND INJECTOR GEOMETRY

In a LJIC configuration, two types of waves are observed – surface waves and column waves. As depicted in Fig. 8, surface waves are observed in the near nozzle region and are a result of the interaction between the crossflow air and injected liquid. As it forms very near the injector exit, exit flow conditions play an important role in its formation, as will be demonstrated later in this section. Column waves are predominantly controlled by the aerodynamic forces from crossflow and have little effect on injector exit conditions. In this section, we will focus on surface waves and try to understand the influence of injector geometry in their formation.

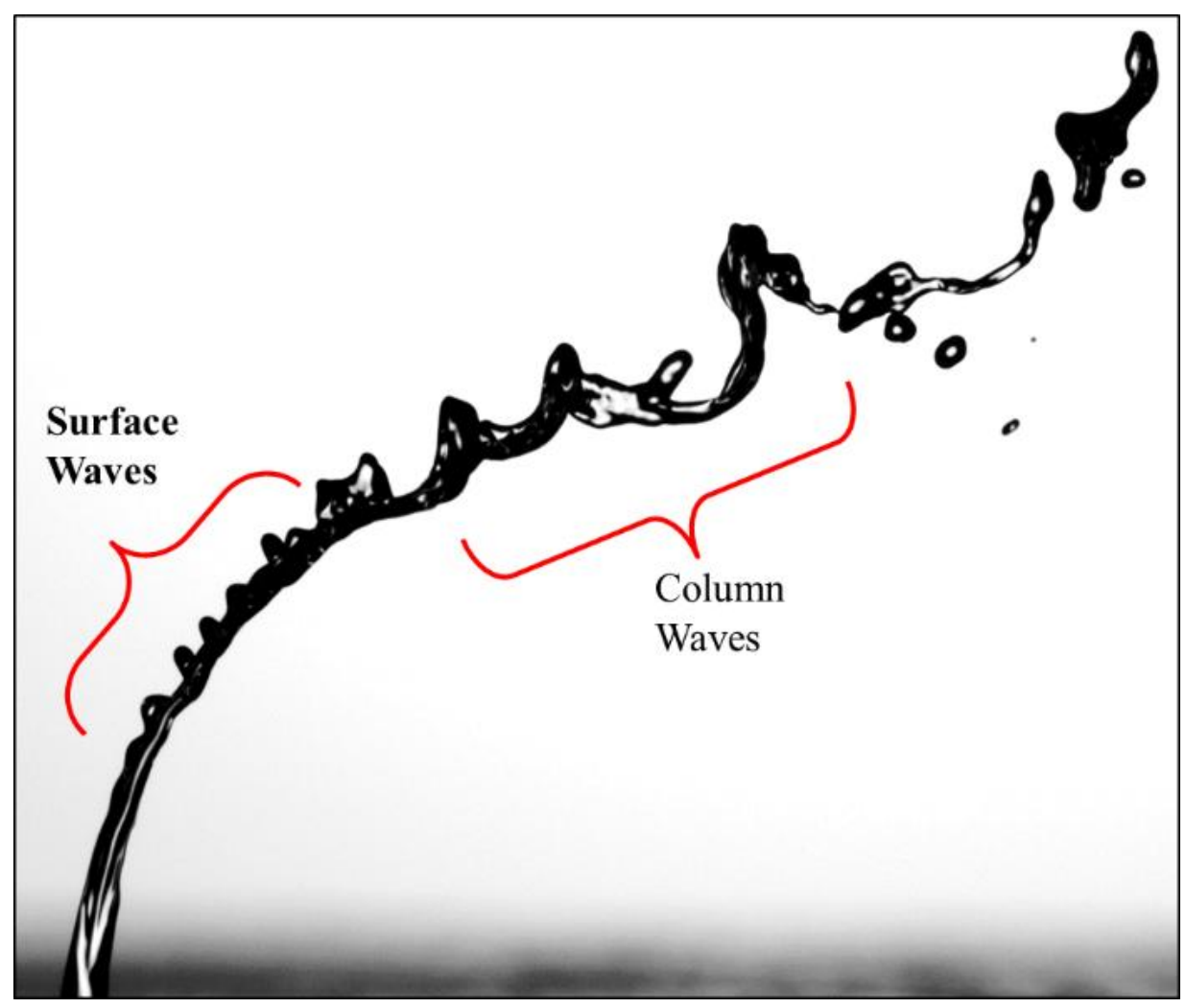

**Fig. 8.** Waves on a liquid jet in crossflow

Surface waves are one of the earliest instabilities that manifest on the liquid jet surface in crossflow. A lighter fluid (air) pushing against a heavier fluid (liquid) is a classic example of Rayleigh-Taylor instability. The same configuration is observed in LJIC. Surface waves on LJIC have been observed experimentally by various researchers [17-20]. There have also been theoretical derivations based on Rayleigh-Taylor instability to estimate the wavelength of the fastest growing (or dominant) wave which is observed in experiments. It is interesting to note that all previous researchers agree that the wavelength of the fastest-growing wave depends only on the operating conditions, particularly the Weber number. The correlations for predicting surface waves are generally of the form:

$$\frac{\lambda}{D} \propto We^{-n} \tag{6}$$

where, $\lambda$ is the wavelength of the wave, $D$ is the jet diameter, $We$ is the Weber number and $n$ is a curve-fit constant. Equation (6) implies that the wavelength should remain fixed for a constant $We$. To examine the effect of injector geometry, particularly L/D, an experimental study was conducted [20] where surface waves were recorded for injectors with different L/D, for the same operating conditions. Some representative results are shown in Fig. 9. As evident, a large difference in wavelength values is observed for the same operating condition and $We$.

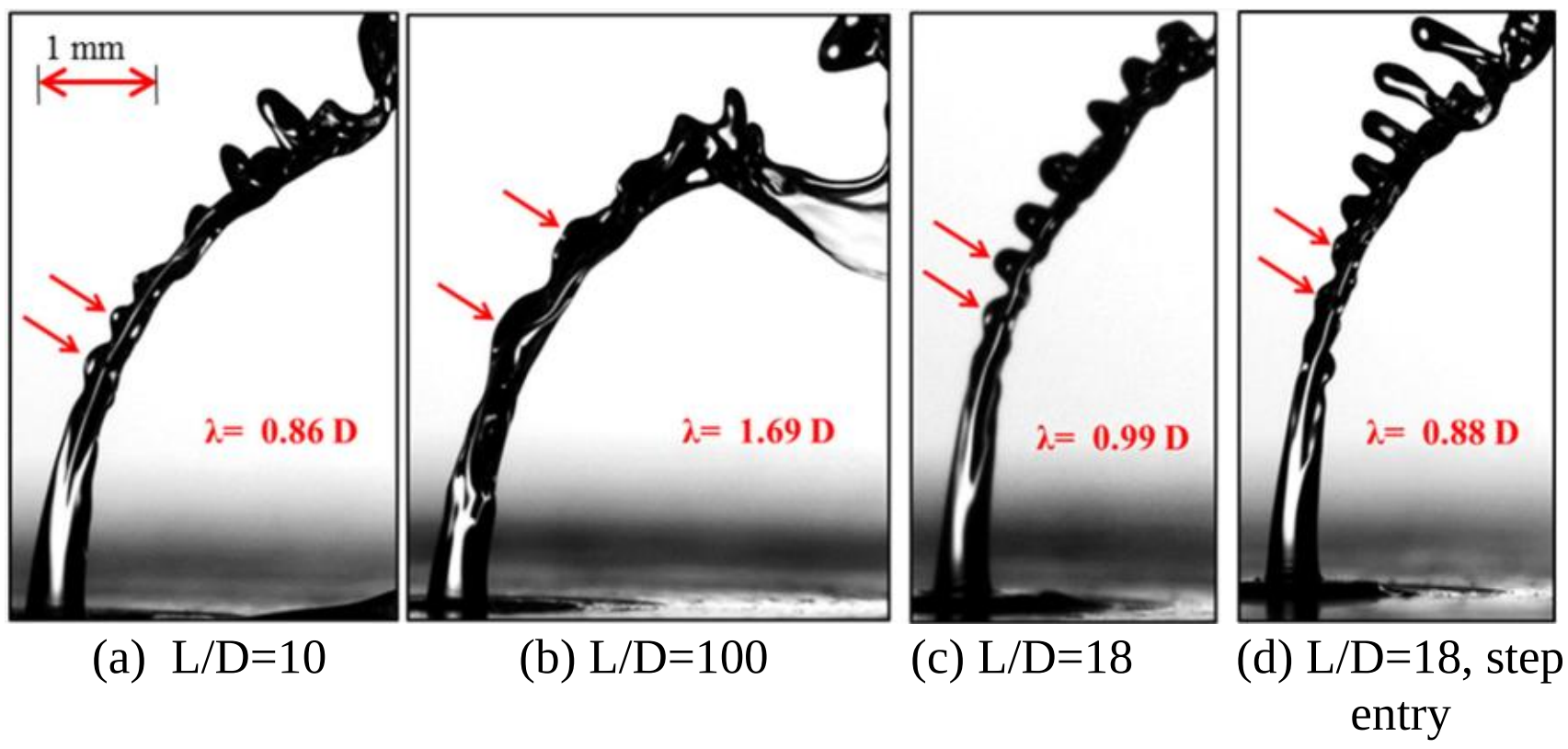

(a) L/D=10 (b) L/D=100 (c) L/D=18 (d) L/D=18, step entry

**Fig. 9.** Wavelength of surface waves for different L/D ratios, and $We$=16 (from Sinha [20])

The wavelength of the surface waves can be estimated using theoretical derivations based on Rayleigh-Taylor instability [20]. The growth rate of a small perturbation at the interface of two fluids with different densities is given by Chandrasekhar [21]:

$$\eta^2 = ak\left[\frac{\rho_l - \rho_g}{\rho_l + \rho_g} - \frac{k^2\sigma}{a(\rho_l + \rho_g)}\right] \tag{7}$$

where $\eta$ is the growth rate of disturbance, $k$ is the wave number, $\rho_l$ and $\rho_g$ are fluid densities, and $a$ is the acceleration. Assuming air density to be negligible in comparison with the liquid density, Eq. (7) can be reduced to:

$$\eta^2 = \left[ak - \frac{k^3\sigma}{\rho_l}\right] \tag{8}$$

A similar form of equation has been used by Ng et al. [19] for column waves, where they have assumed that the whole liquid column deforms due to aerodynamic forces. This is an appropriate assumption for column waves, but not for surface waves. As observed in Fig. 9, the entire liquid column is not affected by surface waves. Assuming an effective thickness *h* participating in the formation of surface waves, which respond to aerodynamic forcing by air crossflow:

$$\frac{1}{2}\rho_g V_g^2 = \rho_l a h \tag{9}$$

From Equations (8) and (9):

$$\eta = \left[\frac{k\rho_g V_g^2}{2\rho_l h} - \frac{k^3\sigma}{\rho_l}\right]^{1/2} \tag{10}$$

Differentiating Eq. (10) with respect to wavenumber (*k*), and equating it to zero to attain the maxima:

$$\left[\frac{\rho_g V_g^2}{2\rho_l h} - \frac{3k^2\sigma}{\rho_l}\right] = 0, \tag{11}$$

or,

$$k = \sqrt{\frac{\rho_g V_g^2}{6\sigma h}}. \tag{12}$$

The wavenumber (*k*) can be expressed in terms of wavelength (λ) as:

$$k = \frac{2\pi}{\lambda}. \tag{13}$$

Hence:

$$\lambda = 2\pi\sqrt{\frac{6\sigma h}{\rho_g V_g^2}}. \tag{14}$$

Non-dimensionalizing by injector diameter $d_j$ , and using the expression of *We* (from Eq. (1)):

$$\frac{\lambda}{D} = \frac{2\pi\sqrt{6}}{\sqrt{We}}\sqrt{\frac{h}{d_j}}. \tag{15}$$

For the near-nozzle region in LJIC, it can be assumed that the jet behaves similarly to the jet in quiescent ambient, and momentum thickness ($\theta$) determines the value of effective thickness (*h*). Hence, the effective thickness is assumed to be expressible in the form:

$$h = \sqrt{\frac{\upsilon L}{V_l}} \tag{16}$$

where L is the injector tube length. By incorporating the injector dimensions in this derivation, there can be a theoretical basis to explain the observation in previous sections where the injector geometry role is observed in surface wave growth and jet breakup . As evident, injectors with the same D and different L have different boundary layer thicknesses, and that influences the breakup behavior. Substituting *h* in Eq. (15):

$$\frac{\lambda}{d_j} = 2\pi\left[\frac{6}{d_j We}\right]^{1/2}\left[\frac{\upsilon L}{U_l}\right]^{1/4} \tag{18}$$

This expression is used to estimate wavelength values for different injectors at various conditions. Comparison of estimated and measured wavelengths from [20] is shown in Fig. 10(a). A line y = x is also superimposed to assess the degree of match between the experimentally measured and predicted values. Further, data on surface wave wavelengths available in the literature [17-20] is compared with the expressions proposed in those papers [17-20] and present derivation [20] in Fig. 10(b). As observed, the present expression (Eq. 18) is more accurate in predicting the data available in literature as compared to other available relations. It also offers a suitable explanation for the dependence of jet breakup behavior on injector geometry

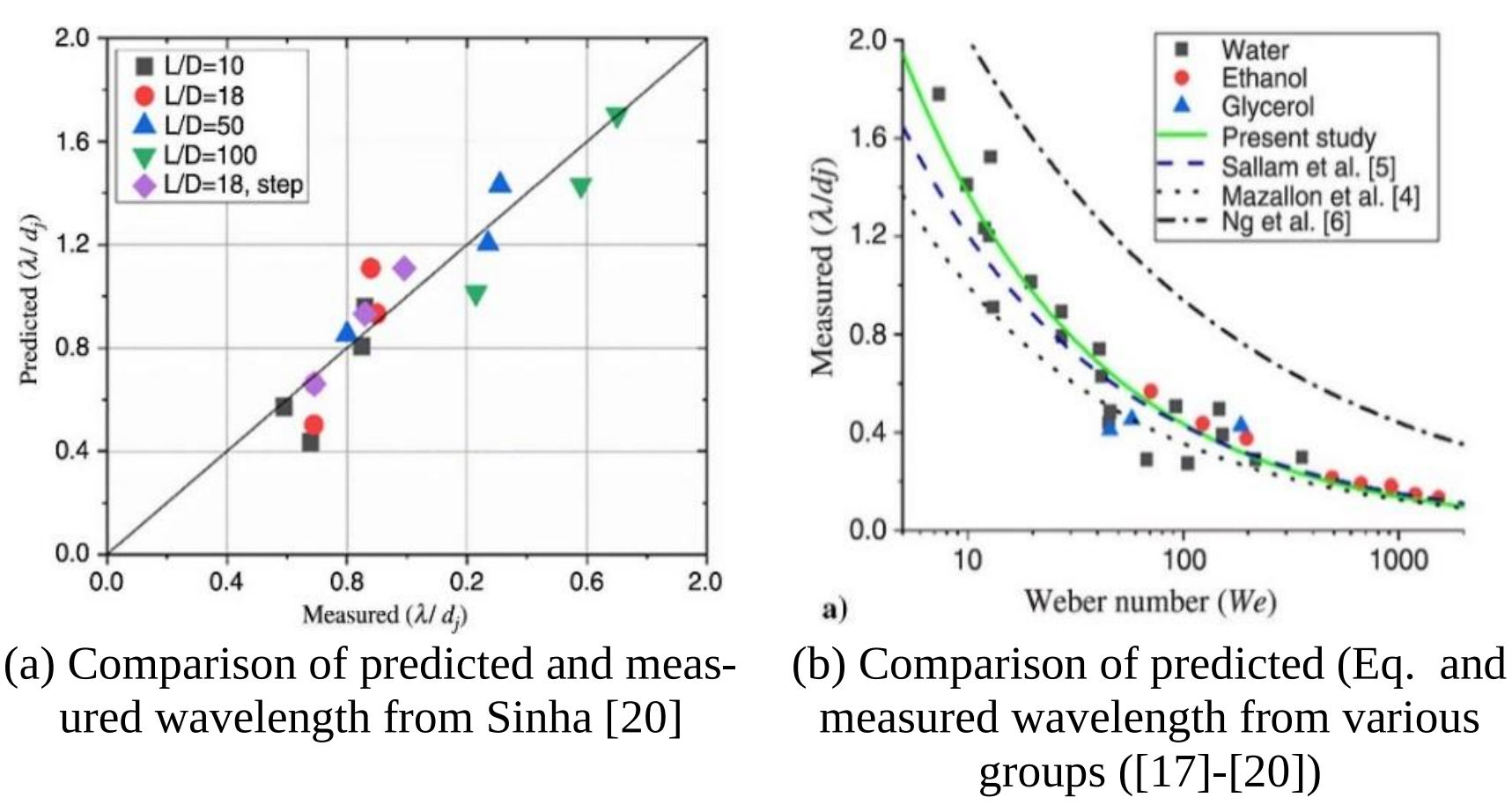

(a) Comparison of predicted and measured wavelength from Sinha [20]

(b) Comparison of predicted (Eq. and measured wavelength from various groups ([17]-[20])

**Fig. 10.** Comparison of wavelength predictions with measured values for surface waves on LJIC

## 5. JET PENETRATION AND TRAJECTORY

Fuel impingement on combustor walls and near-wall combustion is detrimental to performance, engine liner life, and emissions. Therefore, for the entire range of operating conditions of an engine, the jet is expected to neither hit the top wall nor be lying too close to the bottom wall. Hence, liquid trajectory in LJIC is a critical parameter from a design perspective, and numerous efforts have been made to accurately predict it. There are theoretical investigations where the jet is assumed to act like a cylinder that is bending in the presence of crossflow. The jet trajectory is predicted by the force balance on the liquid jet. However, the problem is not so simple. A liquid jet in LJIC configuration is not a cylinder. It gets flattened by the crossflow [22] and consequently the drag increases. Eventually, it will break and produce droplets and ligaments which further complicate the analysis.

However, detailed derivations are presented in the literature with some simplified assumptions which work reasonably well. Applying force balance might be more accurate on the droplet as it is not attached and might not deform as much as the jet, especially if the droplet is small. Another approach to predict jet trajectory is to use the available experimental data and derive a correlation for trajectory. There is a plethora of available literature with such trajectory correlations. No [23] has presented a review of such studies. The jet trajectory correlations are usually expressed in power law form as [22]:

$$\frac{y}{D} = A \cdot q^{\alpha} \cdot \left(\frac{x}{D}\right)^{\beta} \tag{19}$$

where A, $\alpha$, and $\beta$ and are correlation constants and $q$ is the momentum ratio. Another popular form is the logarithmic form:

$$\frac{y}{D} = A \cdot q^{\alpha} \cdot \ln\left[1 + \beta \left(\frac{x}{D}\right)\right] \tag{20}$$

Generally, momentum ratio ($q$) is the only parameter used in most of these correlations, implying that jet penetration is governed by $q$ only. However, some correlations also use other parameters, like $We$. The correlations using $We$ are both in power form and logarithmic form. Power form correlations can be expressed as:

$$\frac{y}{D} = A \cdot q^{\alpha} \cdot \left(\frac{x}{D}\right)^{\beta} \quad We^{\gamma} \tag{21}$$

$\gamma$ is another model constant. Logarithmic form of correlation with Weber number dependence can be written in the form:

$$\frac{y}{D} = A \cdot q^{\alpha} \cdot \left[B \ \ln\left(\frac{x}{D}\right) + C\right] \ We^{\beta} \tag{22}$$

where A, B, C, $\alpha$, and $\beta$ and are correlation constants. Weber number determines the breakup mode and the degree of atomization, which will govern whether the jet is penetrating as a column or a group of droplets. Hence, Weber number dependence is required in these equations. However, the initial jet momentum relative to the airflow momentum will determine the penetration of the liquid jet, hence q has a dominant role, even for equations including Weber number. There are also some other correlations with other parameters, like viscosity ratio, etc. Momentum ratio-based equations are widely accepted and are commonly used. It is observed that there are numerous correlations for a seemingly simple configuration of LJIC. And their predictions vary greatly, even for the same equation form, as the constants are very different [22]. There are many reasons for this scatter.

First, let us understand how this correlation is derived. A large number of instantaneous images are captured to account for the unsteady nature of the flow. Then averaged images are used to determine the trajectory points. Several such experiments for a range of conditions are used with curve-fitting algorithms to find the model constants. Now, each study has a different range of conditions, and equations derived for a particular study might work best for the range of conditions used in that study. Moreover, the curve-fitting algorithms or multi-variable regressions only give the model constants, the form of the equation is to be pre-decided by the user. This will also cause discrepancy, as the preference of parameters becomes subjective, and will also depend on the range of conditions studied. Discrepancy may also arise due to different experimental techniques used, different methods of image processing, etc. One common issue that is often overlooked is the role of injector geometry. In previous sections, it was shown that the role of injector geometry was neglected in surface wave predictions and each research group had a different correlation for wavelength prediction. However, with the inclusion of geometry effects, a good match was found for available data in the literature. Similarly, for jet trajectory prediction, Prakash et al. [14] have proposed a correlation that incorporates injector (L/D) in the correlation:

$$\frac{y}{D} = 1.831 \cdot q^{0.428} \cdot \left(\frac{x}{D}\right)^{0.53} \cdot \left(\frac{L}{D}\right)^{-0.0216} \qquad (23)$$

This equation works well with different injectors. Model predictions are compared with data points from experiments, and a reasonably good match is found [14]. Still, this appears to be an area where more research is required to gain a deeper understanding of liquid jet breakup and trajectory predictions.

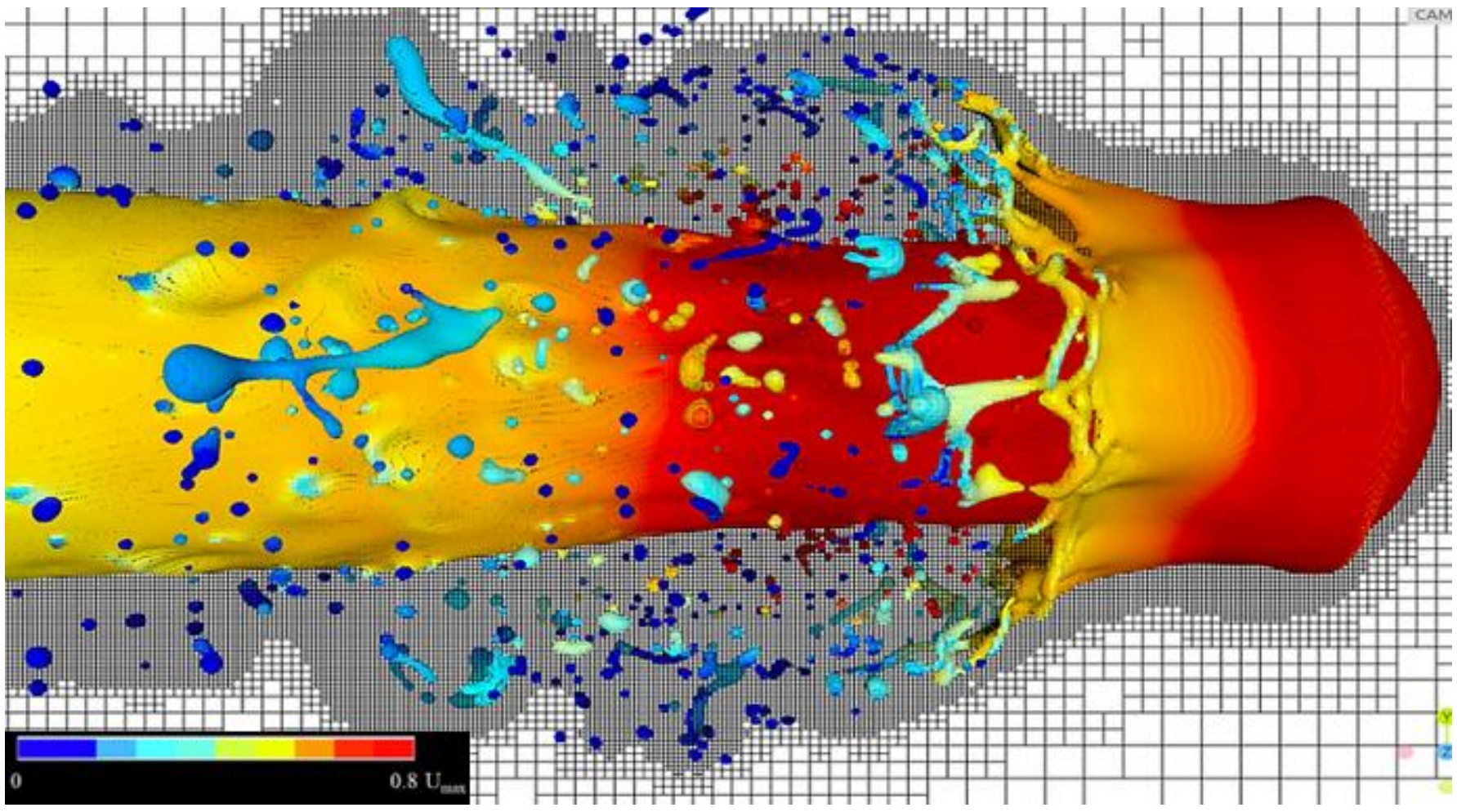

**Fig. 11. DNS of** Liquid jet breakup – highlighting adaptive meshing.

## 6. COMPUTATIONAL MODELING OF LJIC

Computational modeling provides a convenient tool that can examine any flow field non-intrusively and simulate extreme conditions that are challenging to attain in laboratory scale experiments. However, the major challenge is to get a model which is accurate and reliable. Modeling a multiphase flow field like LJIC is a daunting task, especially due to the sudden jump in density across the interface. The common fuels used is typically around 600 times denser than the atmospheric air. This poses several challenges for numerical models and they often blow up. However, with the progress in numerical algorithms, several methods have been developed to compute the large density ratio multiphase flows. Volume of Fluid (VoF), Level set methods, etc. are generally used for LJIC. A comprehensive review of numerical methods can be found in some recent works [24, 25]. Other than the interface capturing technique, the type of simulation will also determine the accuracy of the results. Generally, LES is not able to accurately capture the jet dynamics and DNS is required for reasonable predictions [25]. Jet morphology and the breakup process are very challenging to model in LES and a DNS will be required for proper understating of the flow physics. DNS of a multiphase flow requires massive computational costs even for very small computational domains. However, with the recent development in efficient algorithms and adaptive mesh methods [27, 28], the future of high-performance computing of LJIC seems to have a promising future. Figure 11 illustrates the adaptive meshing strategy for DNS of a liquid jet. Note how finer mesh is created near the interface and droplet clusters and coarser mesh is present in regions away from liquid surface.

Hermann [29] has carried out one of the earliest DNS studies of LJIC configuration and has explored the effect of ambient pressure on the jet structure. Column and surface breakup are observed in those simulations. It is interesting to note that for capturing the physical process correctly, flow inside injector geometry is also simulated. Behzad et al. [30] have also carried out DNS of LJIC. They can observe surface waves and shear striping from the jet, which is a challenging physical aspect to capture. A typical plot of DNS results of LJIC is presented in Fig. 12. The liquid surface is colored by liquid axial velocity. As evident, ligament and drop formation are captured which exemplify the strengths of DNS study. These DNS results look promising and more such studies are expected to reveal the breakup mechanism and dynamics of LJIC configuration.

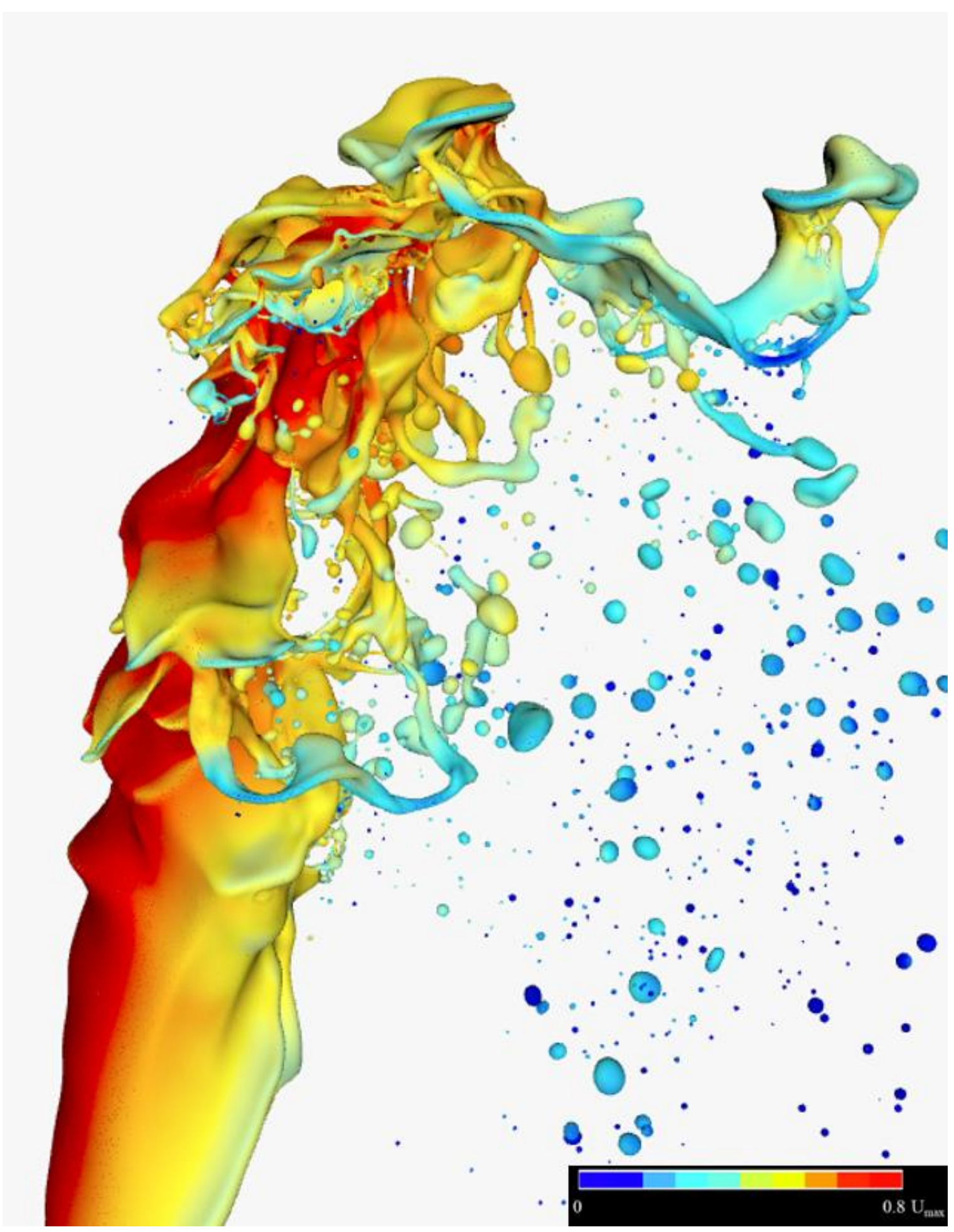

**Fig. 12. DNS of** Liquid jet in crossflow capturing ligament and droplet formation.

## 7. SUMMARY

This chapter presents an overview of the Liquid jet in crossflow (LJIC) configuration. The idea is to elaborate on the physical processes involved with high-resolution experimental and computational images. In the first section, the topic is introduced and its significance is highlighted. In the second section, various LJIC breakup modes are identified and explained in detail. High-resolution experimental images are presented to gain an understanding of breakup modes and breakup regimes. Important parameters are identified and listed. Special attention is given to injector geometry which seems to influence breakup but is often overlooked. Various experiments are conducted to identify the role of injector geometry. Surface waves are also explained, and their formation is studied analytically, with the derivation of a relation based on injector geometry. Further jet trajectory and penetration are discussed. Different correlation forms and significant parameters affecting them are listed. The last section gives a brief overview of advances in computational modelling and inherent challenges. Overall LJIC is an important area of research owing to its rich physics and relevance to practical applications. Much more research is expected in the coming time to dive deeper into the breakup mechanisms and their underlying physics.